\documentclass[%
 reprint,nofootinbib,superscriptaddress,
 amsmath,amssymb,
 aps,
pre,
]{revtex4-1}

\usepackage{graphicx} % Include figure files
\usepackage{bm}
\usepackage{txfonts}
\usepackage{mathrsfs}
\usepackage{amsmath}
\usepackage{xcolor}
\usepackage[normalem]{ulem}
\usepackage{subcaption} 

\def\b0{{\mathbf 0}}

\def\b0{{\mathbf 0}}

\def\eps{\epsilon}

\def\beq{\begin{equation}}
\def\eeq{\end{equation}}
\def\beqa\begin{eqnarray}
\def\eeqa{\end{eqnarray}}

\begin{document}

%\title{Nature of the superfluid quantum phase transition in imbalanced Fermi mixtures} 
%\title{Stabilizing the Fulde-Ferrell-Larkin-Ovchinnikov states with anisotropies: \\role of the Lifshitz point} 
\title{$\mathbb{Z}_4$-symmetric perturbations to the XY model from functional renormalization}
\author{Andrzej Chlebicki}
\affiliation{Institute of Theoretical Physics, Faculty of Physics, University of Warsaw, Pasteura 5, 02-093 Warsaw, Poland}
%\affiliation{Max-Planck-Institute for Solid State Research, Heisenbergstr.\ 1, D-70569 Stuttgart, Germany}
\author{Carlos A. S\'anchez-Villalobos }
\affiliation{Instituto de F\'isica, Facultad de Ingenier\'ia, Universidad de la Rep\'ublica, J.H.y Reissig 565, 11300 Montevideo, Uruguay}
\author{Pawel Jakubczyk }
\affiliation{Institute  of Theoretical Physics, Faculty of Physics, University of Warsaw, Pasteura 5, 02-093 Warsaw, Poland}
\author{Nicol\'as Wschebor}
\affiliation{Instituto de F\'isica, Facultad de Ingenier\'ia, Universidad de la Rep\'ublica, J.H.y Reissig 565, 11300 Montevideo, Uruguay} 
\date{\today}
\begin{abstract}
We employ the second order of the derivative expansion of the nonperturbative renormalization group to study cubic ($\mathbb{Z}_4$-symmetric) perturbations to the classical $XY$ model in dimensionality $d\in [2,4]$. In $d=3$ we provide accurate estimates of the eigenvalue $y_4$ corresponding to the leading irrelevant perturbation and follow the evolution of the physical picture upon reducing spatial dimensionality from $d=3$ towards $d=2$, where we approximately recover the onset of the Kosterlitz-Thouless physics. We analyze the interplay between the leading irrelevant eigenvalues related to $O(2)$-symmetric and $\mathbb{Z}_4$-symmetric perturbations and their approximate collapse for $d\to 2$.  We compare and discuss different implementations of the derivative expansion in cases involving one and two invariants of the corresponding symmetry group. 

\end{abstract}

\pacs{}

\maketitle

%%%%%%%%%%%%%%%%%%%%%%%%%%%%%%%%%%%%%%%%%
\section{Introduction}
Cubic symmetry-breaking fields occur in lattice systems, where the ions are arranged into a crystal structure characterized by a (hyper)cubic symmetry. Prominent examples are the  $XY$ or Heisenberg magnets \cite{Zinn_Justin_small, Cardy_book}. From the point of view of the renormalization group (RG) the cubic ($\mathbb{Z}_4$-symmetric in the case of $XY$ magnets) perturbation was, to our knowledge, first addressed by Aharony \cite{Aharony_1973} within the $\epsilon=4-d$ expansion. Such a perturbation can be  relevant, marginal, or irrelevant at the $O(N)$-symmetric RG fixed point, depending on spatial dimensionality $d$ and the number of field components $N$ \cite{Carmona_2000}. 
 For low dimensionalities [$d < d_c(N)$], the $O(N)$ symmetric fixed point is unstable with respect to the cubic perturbation and the phase transition is controlled by a different fixed point displaying only cubic symmetry. For $d$  large [$d>d_c(N)$] the cubic perturbation is irrelevant in the sense of the renormalization group. However, in this situation, the cubic anisotropy constitutes a so called dangerously irrelevant operator; the presence of such a perturbation, whatever small, modifies the critical exponents governing the dominant scaling close to the transition. In the present case, the physical reason for this is that the discrete anisotropies gap the Goldstone mode in the ordered phase and give rise to an additional length scale, divergent at the phase transition. This leads in particular to the dependence of the critical exponents on the side from which the phase transition is approached \cite{Nelson_1976, Leonard_2015}. For the $XY$ model ($N=2$) the cubic perturbation is dangerously irrelevant for $d=3$ as well as for $d=4-\epsilon$, and marginal for $d=2$. A natural expectation  (confirmed by the results of the present paper) is that it remains dangerously irrelevant for $d\in ]2,4[$.

Additionally, the cubic perturbation induces significant corrections to scaling due to very small value of the corresponding scaling dimension. Indeed, in the three-dimensional $O(2)$ model the $\mathbb{Z}_4$ anisotropy is irrelevant, but the associated scaling dimension  takes the value $|y_4| \approx 0.1$ \cite{Carmona_2000, Hasenbusch_2011, Okubo_2015}, whereas the first isotropic correction is characterized by the exponent $\omega \approx 0.8$ \cite{Pelissetto_2002}. 

For a time it remained controversial, whether the cubic anisotropy is relevant at the Heisenberg fixed point in $d=3$ [i. e. at which side of the line $d_c(N)$ the point $(d=3, N=3)$ is situated]. This issue seems to have very recently been completely settled \cite{Chester_2021} by the numerical conformal bootstrap approach, which (numerically) proves the weakly relevant character of the cubic anisotropy at the $O(3)$ fixed point. This is in agreement with the predictions of the perturbative RG up to six loop order \cite{Carmona_2000}, but not earlier, lower level calculations. The precise shape of the line $d_c(N)$ for $d\in [2,4]$ is in general not easy to compute reliably. It is known \cite{ Jose_1977, Delfino_2019} to pass through the points $(N,d)=(2,2)$, as well as close vicinity of $(N,d)=(3,3)$ and $(N,d)=(4,4)$ \cite{Kleinert_1995}.

%The point $(d=2,N=2)$ lies on the line $d_c(N)$ and in dimensionality $d=2$ the cubic anisotropy is marginal at the Kosterlitz-Thouless fixed point. It acts to stabilize the long-range-ordered phase and yields the interesting  physical situation, where the critical singularities are controlled by an entire line of fixed points, such that the corresponding critical indices are not universal. The $XY$ model with $\mathbb{Z}_4$ anisotropies is of high relevance in many experimental contexts (see Ref. \dots  for a useful review). 

The point $(d=2,N=2)$ lies on the line $d_c(N)$ and in dimensionality $d=2$ the cubic perturbation stabilizes the long-range order destroying the Kosterlitz-Thouless (KT) phase \cite{Jose_1977}. Exactly at the temperature of the KT transition, the corresponding coupling $h_4$ becomes marginal, which leads to formation of a line of fixed points intersecting with the KT line of fixed points at the temperature of the KT transition and $h_4=0$. The continuous transition controlled by the anisotropic line of fixed points is characterized by nonuniversal critical exponents \cite{Jose_1977, Delfino_2019}. The $XY$ model with $\mathbb{Z}_4$-symmetric anisotropies is of high relevance in many experimental contexts (see Ref.~\onlinecite{Taroni_2008}  for a useful review).

In the present paper we address the cubic perturbations to the $XY$ model from the point of view of nonperturbative renormalization group (NPRG). In very recent years there has been remarkable progress \cite{Dupuis_2021} in implementing this methodology for high-precision calculations of the critical properties of the isotropic $O(N)$ models in $d=3$. The NPRG approach also continues to deliver new insights inaccessible within the perturbative paradigms. Of our primary interest in this work is the leading irrelevant perturbation to the $O(2)$-symmetric fixed point and its evolution upon continuously evolving dimensionality from three towards two, where the corresponding eigenvalue vanishes. The NPRG constitutes a unique approach potent to capture the entire range of $d\in [2,4]$. This however requires a substantial methodological advancement as compared to the pure $O(N)$-symmetric systems, since the effective action depends on two invariants and the analysis must start off from a two-dimensional field space. As we demonstrate in this article, NPRG is able to reach precise and accurate results in $d=3$ together with a proper qualitative description of the $\mathbb{Z}_4$ model down to $d=2$.

The outline of the paper is as follows: In Sec. II we present the framework of functional RG and the derivative expansion (DE) scheme employed in our analysis. In Sec. III we lay down the results for the three-dimensional situation (Sec.~IIIA) and reducing dimensionality towards two (Sec.~IIIB). We compare the results obtained from different truncations of the DE applied by us. In Sec.~IV we provide a discussion of different implementations of the DE (the so-called ``ansatz" and ``strict" versions). We point out the differences, which appear  relevant for $d$ close to two. Sec. V contains a summary and conclusions.

\section{Model and methodology}
We consider the canonical $O(2)$-symmetric $\phi^4$ model perturbed by a $\mathbb{Z}_4$-symmetric operator. A prototype effective action (or Hamiltonian) analyzed by us takes the form:
\begin{equation} 
\label{bare_action}
    \mathcal{S} = \int d^dx \left[\frac{1}{2} \left(\nabla \bm \phi(x)\right)^2+ \frac{\lambda}{8}\left(\bm \phi(x)^2 - \phi_0^2\right)^2 + \frac{h_4}{2} \phi_1^2 \phi_2^2\right], 
\end{equation}
where $\lambda$ is the standard quartic coupling, $\phi=(\phi_1,\phi_2)$ represents the order parameter field and $h_4$ describes the strength of the anisotropic perturbation. The above action is defined on a coarse-grained length scale $1/\Lambda$ and serves as an effective model representing the same universality class as more microscopic models, typically defined on a lattice. An example of such a microscopic model, introduced in Ref.~\cite{Jose_1977}, is given by the Hamiltonian 
\begin{equation}
\mathcal{H}=-\tilde{J}\sum_{\langle i,j\rangle}\cos(\theta_i-\theta_j)+\tilde{h_4}\sum_i \cos(4\theta_i)\;,
\end{equation}
where $i$ labels the sites of a lattice, $\sum_{\langle i,j\rangle}$ denotes summation over nearest neighbor pairs, while $\theta_i$ describes the orientation of a 2-dimensional spin-like degree of freedom associated to each lattice site, relative to a given axis. The action given by Eq.~(\ref{bare_action}) is invariant under the two independent symmetry transformations: 
\beq
\phi_1 \longleftrightarrow -\phi_1\;,\;\;\;\;\; \phi_1 \longleftrightarrow \phi_2\;.
\eeq 
  It is well established that at the Wilson-Fisher fixed point in $d=3$, the quantity $h_4$ couples to a dangerously irrelevant perturbation characterized by a relatively small  eigenvalue $y_4\approx -0.1$ \cite{Carmona_2000, Hasenbusch_2011, Okubo_2015}. In particular, $|y_4|$ is small as compared to the absolute value of the leading irrelevant eigenvalue related to an $O(2)$-symmetric perturbation  $\omega\approx 0.8$ \cite{Pelissetto_2002}. On the other hand, the quantity $y_4$ vanishes upon reducing dimensionality towards $d=2$ \cite{Jose_1977}. As already mentioned, the presence of the dangerously irrelevant perturbation gives rise to the differences between the  critical exponents depending on the side from which the transition is approached. The magnitudes of these differences are controlled by the value of $y_4$. In particular, one finds \cite{Leonard_2015}: 
\begin{align}
\label{full_scaling}
&\nu' = \nu (1 + |y_4|/2)\;, \nonumber\\
&\gamma^+ = \nu(2-\eta)\;, \\
&\gamma_T = \gamma^+ + \nu |y_4|\;, \nonumber\\
&\gamma_L = \gamma^+ + \nu |y_4| \frac{4-d}{2}\;, \nonumber
\end{align}
where $\gamma_T$, $\gamma_L$ are the critical exponents for the transverse and longitudinal susceptibilities in the low-temperature phase, $\gamma^+$ controls the susceptibility divergence when the critical point is approached from the high-temperature phase,  while $\eta$ is the anomalous dimension. The quantities $\nu$ and $\nu'$ denote the critical exponents for the longitudinal and transverse correlation lengths. Eq.~(\ref{full_scaling}) is valid for $y_4<0$, i.e. whenever the anisotropy is irrelevant and therefore is expected to hold for $d\in ]2,4[$ for them model defined by Eq.~(\ref{bare_action}). The modification of the scaling laws as compared to the pure $O(N)$ case arises due to the presence of two large length scales associated to the two directions in the field space. These lengths are determined by the RG scales where the RG flow diverges from the $XY$ and the low-temperature fixed points respectively \cite{Leonard_2015}. In the absence of anisotropies, the latter scale is always infinite. In $d=4$, the coupling $h_4$ becomes marginal and it can be treated perturbatively when $d=4-\epsilon$ and $\epsilon\ll 1$. 

There are not many independent accurate estimates of the quantity $y_4$ in dimensionality $d=3$ (see Sec.IIIA for more details) and the earlier NPRG predictions yield values which disagree with those obtained from perturbative RG by a factor larger than 2 \cite{Leonard_2015, Chlebicki_2019}. In addition, we are not aware of any earlier implemented theoretical framework capable of capturing the continuous interpolation of $y_4$ between $d=3$ and $d=2$ and delivering a comparison with leading eigenvalues related to $O(2)$-symmetric perturbations. The present paper aims at filling this gap. 
 
Independent of the above, the model defined by Eq.~(\ref{bare_action}) displays rich and  interesting crossover behavior related to the interplay between three different fixed points in $d=3$. On the other hand, in $d=2$ it exhibits \cite{Jose_1977} a line of fixed points extending towards positive and negative values of $h_4$, merging with the KT line at $h_4=0$. For a qualitative resolution of this picture within a simple truncation of NPRG, see Ref.~\onlinecite{Chlebicki_2019}. For large anisotropies ($|h_4|>\lambda$) the system displays a fluctuation-induced first order transition \cite{Carmona_2000}.

We address the model defined  by Eq.~(\ref{bare_action}) within the one-particle-irreducible variant of NPRG, relying on the Wetterich equation \cite{Wetterich_1993, Morris_1993, Ellwanger_1993} 
\begin{equation}
    k \partial_k \Gamma_k[\phi] = \frac{1}{2} \mathrm{Tr} \left[k \partial_k R_k \left(\Gamma_k^{(2)}[\phi]+ R_k\right)^{-1}\right]\;.
    \label{Wetterich_eq}
\end{equation}
This describes the renormalization flow of the effective average action $\Gamma_k[\phi]$ - a scale-dependent functional of the order parameter field $\phi$. The flow is initiated at a microscopic scale ($k=\Lambda$) with the effective action equal to the microscopic action $\Gamma_{\Lambda} = \mathcal{S}$. As the scale $k$ decreases, the effective action smoothly evolves towards the Gibbs free energy $\mathcal{F}$, which is reached in the limit of vanishing cutoff scale $\Gamma_{k\rightarrow 0} = \mathcal{F}$. In Eq.~(\ref{Wetterich_eq}) the quantity $\Gamma_k^{(2)}[\phi]$ denotes the second (functional) derivative of $\Gamma_k[\phi]$ with respect to the field $\phi$, $R_k$ is the momentum cutoff function added to the inverse propagator, which damps fluctuations with momentum $q$ below the scale $k$, and the trace $\mathrm{Tr}$ sums over momenta and the field components. In the present context we specify $N=2$ so that the field $\phi=(\phi_1, \phi_2)$ contains two cartesian components.  The framework based on the Wetterich equation has since long been successfully applied in a diversity of physical contexts [see Refs.~\onlinecite{Berges_2002, Delamotte_2004, Pawlowski_2007, Kopietz_book, Metzner_2012, Gies_2012,  Dupuis_2021} for reviews]. However, only very recently has it been promoted to a computational tool capable of delivering high-precision numerical estimates of the universal properties of the $O(N)$ models in $d=3$ \cite{Balog_2019, Balog_2020, Polsi_2020, Polsi_2021}, which appear of accuracy comparable to the best Monte-Carlo simulations and conformal bootstrap estimates. This precision is associated with the existence of a small parameter (which  is estimated to lie between 1/9 and 1/4) that controls the size of successive orders of the most commonly employed approximation in this context, the derivative expansion.
 Within the context of pure $O(N)$ models, the NPRG methodology has also recently revealed a new family of perturbatively inaccessible fixed points \cite{Yabunaka_2017, Yabunaka_2018} and reopened the question concerning the analyticity of the critical exponents \cite{Chlebicki_2021} when viewed as functions of $d$ and $N$. These findings may appear quite unexpected bearing in mind that the $O(N)$ models are among the most broadly recognized and well-studied systems exhibiting critical behavior. In other statistical physics contexts, the NPRG approach led, {\it inter alia} to the resolution of the puzzle concerning dimensional reduction and its breakdown in the random field Ising model \cite{Tissier_2011} as well as to important insights into the Kardar-Parisi-Zhang problem in dimensionalities $d>1$ \cite{Canet_2010, Canet_2011}. It also constitutes a remarkably suitable tool for addressing physical situations involving rich crossover behavior (see e.g. Refs.~\onlinecite{Leonard_2015, Strack_2009, Jakubczyk_2010, Lammers_2016, Debelhoir_2016, Rancon_2016,  Rancon_2017, Zdybel_2020}).      

 We point out that for the pure $O(N)$ models the framework resting upon Eq.~(\ref{Wetterich_eq}) is capable of addressing the entire range of physically relevant values of $d$ and $N$ with a single set of flow equations derived within the DE scheme. 

The DE constitutes a fertile approximation scheme, which allows for casting Eq.~(\ref{Wetterich_eq}) onto a closed set of numerically tractable partial (integro)differential equations. It amounts to expanding the effective action $\Gamma_k$ in powers of gradient operators around the homogeneous field configuration, but at the same time refraining from any expansion in the field $\phi$. For the isotropic $O(N)$ model in $d=3$, the DE was implemented up to order $\partial^4$ \cite{Polsi_2020, Polsi_2021, Peli_2021}, and for the Ising case, even up to order $\partial^6$ \cite{Balog_2019}. For an earlier study of the Ising universality class at order $\partial^4$ see Ref.~\onlinecite{Canet_2003}. 

In the present work, we employ the second-order derivative expansion (DE2) retaining all the symmetry-allowed terms of order up to $\mathcal{O}(\partial^2)$. In the isotropic model ($h_4=0$), the effective action is parameterized by three flowing functions of the $O(2)$ invariant $\rho = \frac{\bm \phi(x)^2}{2}$, with the ansatz: 
\begin{equation}
 %\Gamma_{\mathrm{ISO}} = \int d^d x \left\lbrace U_k(\rho) + \frac{Z_k(\rho)- \rho Y_k(\rho)}{2} \left(\nabla \phi\right)^2  + \frac{Y_k(\rho)}{2} \left(\nabla \rho\right)^2 \right\rbrace
    \Gamma_{\mathrm{ISO}} = \int d^dx \left\{ U_k(\rho) + \frac{Z_k^{\textrm{ISO}}(\rho)}{2} \left(\nabla \bm \phi\right)^2+\frac{Y_k(\rho)}{2}  (\nabla \rho)^2 \right\}\;. 
\label{Ansatz_iso}    
\end{equation} 
By plugging Eq.~(\ref{Ansatz_iso}) into Eq.~(\ref{Wetterich_eq}) and evaluating at a uniform field configuration, one may project the flow of $\partial_k \Gamma_k[\phi]$ onto three partial differential equations governing the flow of the set of functions 
\beq 
\mathcal{F}^{\mathrm{ISO}}_k(\rho) = \{U_k(\rho), Z_k^{\textrm{ISO}}(\rho), Y_k(\rho)\}\;, 
\eeq 
 which constitutes the basis for the numerical analysis. This still poses a relatively complex enterprise involving numerical solutions of complicated nonlinear partial (integro)differential equations. Further simplifications  are often possible [see e.g. Refs.~\onlinecite{Ballhausen_2004, Dupuis_2011, Codello_2013, Rancon_2014, Codello_2015, Borhardt_2016, Millan_2021}] for instance by disregarding the flow of $Z_k^{\textrm{ISO}}(\rho)$, $Y_k(\rho)$ - the so-called local potential approximation (LPA); by downgrading  $Z_k^{\textrm{ISO}}(\rho)$, $Y_k(\rho)$ to $Z_k^{\textrm{ISO}}$ and $Y_k$ (i.e. treating them as flowing, but $\rho$-independent couplings) - the so-called LPA' approximation; or by performing expansion of the flowing functions in $\rho$. This certainly occurs at the cost of accuracy of the results and poor (or disregarded) error control. 
 
 For the present case involving the $\mathbb{Z}_4$ anisotropy, the system is not $O(2)$-symmetric, and we promote the DE2 ansatz to the following form: 
\begin{align}
    \Gamma_{\mathrm{F}} =& \int d^dx \Bigg\{U_k(\rho, \theta) + \frac{Z_k(\rho,\theta)}{2} \left(\nabla \bm \phi\right)^2+T_k(\rho, \theta) \phi_1 \phi_2 \nabla \phi_1 \nabla \phi_2 \nonumber \\
    &+ \left(\phi_1^2-\phi_2^2\right)\frac{W_k(\rho,\theta)}{4} \left(\left(\nabla \phi_1\right)^2-\left(\nabla \phi_2\right)^2\right) \Bigg\}\;, 
    \label{double_functional}
\end{align} 
where 
\begin{align}
\phi_1 =\sqrt{2\rho}\cos\theta \;\,\\
\phi_2 =\sqrt{2\rho}\sin\theta\;.
%\theta = \arctan(\phi_1/\phi_2)\;. 
\end{align} 
and the subscript F in  $\Gamma_{\mathrm{F}}$ stands for `full'. The functions $U_k(\rho, \theta)$, $Z_k(\rho,\theta)$, $T_k(\rho, \theta)$, and $W_k(\rho,\theta)$ are periodic in $\theta$ with period $\frac{\pi}{2}$ and even upon reflection with respect to $\frac{\pi}{4}$. 
The above ansatz is the most general $\mathbb{Z}_4$-symmetric form retaining derivatives up to order $\partial^2$ (i.e. of order $\sim{q}^2$ when written in momentum space). 
%Note that the $\mathrm O(2)$-symmetric term $(\nabla \rho)^2$ is here replaced by the two $\mathbb{Z}_4$-symmetric terms $\phi_1 \phi_2 \nabla \phi_1 \nabla \phi_2$ and $\left(\phi_1^2-\phi_2^2\right) \left(\left(\nabla \phi_1\right)^2-\left(\nabla \phi_2\right)^2\right)$. 
The severe increase of complexity of the problem as compared to the $O(2)$-symmetric case is evident: the flowing functions depend now on two field variables [($\phi_1$, $\phi_2$) or  ($\rho$, $\theta$)], and, in the numerical treatment to follow, must in general be considered on a two-dimensional field grid. The $O(2)$ symmetry is restored in $\Gamma_{\mathrm{F}}$ when all the parameterizing functions are $\theta$ independent and additionally %$W\equiv T$. 
\begin{align}
&T_k(\rho, \theta) \rightarrow Y_k(\rho) \\
&W_k(\rho, \theta) \rightarrow Y_k(\rho) \\ 
&Z_k(\rho, \theta) \rightarrow Z_k^{\textrm{ISO}}(\rho) + \rho Y_k(\rho)\;.
\end{align}

It is important to note, that the integrand in the effective action is a smooth function of the field components $\phi_1$ and $\phi_2$, which imposes constraints. In the polar parametrization via $\rho$ and $\theta$, the continuity of $\Gamma$ implies that $\left.\partial_\theta \mathcal{F}\right|_{\rho=0} = 0$ - the parameterizing functions are $\theta$ independent at the origin. 

We point out here that the previous NPRG studies of discrete anisotropies \cite{Tetradis_1998, Tissier_2002, Leonard_2015, Chlebicki_2019} implemented field expansions (in both $\rho$ and the $\mathbb{Z}_4$ invariant $\tau=\frac{\phi_1^2\phi_2^2}{2}$) and did not systematically account for the momentum dependencies (for example by disregarding the quantities $T_k$ and $W_k$). They largely underestimate the value of $y_4$ in $d=3$, while in $d=2$ they do not fully capture the KT line of fixed points and the marginal character of $y_4$, but instead yield a positive value of $y_4$ for $d\lesssim 2.5$. Note also  that, at $h_4=0$, for the KT transition as well as  the corresponding low-$T$ phase  to be well captured within the present NPRG framework \cite{Graeter_1995, Gersdorff_2001, Jakubczyk_2014, Jakubczyk_2016, Jakubczyk_2017_2, Defenu_2017, Krieg_2017, Rancon_2017}, it is necessary to retain the field dependencies.   

In the analysis to follow, we shall pursue two complementary paths. In the first approach, we will address the parametrization defined by Eq.~(\ref{double_functional}) directly, which leads to flow equations for the functions 
\beq
 \mathcal{F}^{\mathrm{F}}_k(\rho, \theta) = \left\{U_k(\rho, \theta), Z_k(\rho, \theta), T_k(\rho, \theta), W_k(\rho,\theta)\right\} \;.
\eeq
We project out the flow equations for $\mathcal{F}^{\mathrm{F}}_k(\rho, \theta)$ by plugging the ansatz of Eq.~(\ref{double_functional}) into the Wetterich equation [Eq.~(\ref{Wetterich_eq})]. To the best of our knowledge, this is the first NPRG study fully systematically implementing a DE2 level truncation where the functions parametrizing the flowing effective action depend in an essential way on two fields and none of the dependencies are expanded. For studies of other physical situations involving the non-truncated effective potential with two invariants, see Refs.~\onlinecite{Tissier_2011_2, Delamotte_2015}.  

In an alternative procedure, we first express the action using the $O(2)$ and $\mathbb{Z}_4$ invariants, $\rho=\frac{1}{2}\phi^2$ and $\tau = \frac{\phi_1^2\phi_2^2}{2}$ respectively [in particular $U_k(\rho, \theta)\to U_k(\rho, \tau)$], and then expand in $\tau$, retaining the functional form of all $\rho$ dependencies, such that the limit $\tau\to 0$ coincides with the well studied isotropic $O(N)$ case. Up to the linear order in the expansion in $\tau$, which we will consider in the analysis to follow, the effective action ansatz in this scheme reads:
\begin{align}
\label{expanded}
    \Gamma_{\mathrm{E}} =& \int d^dx \Bigg\{U_k(\rho) + \tau U^1_k(\rho) + \frac{Z_{\sigma,k}(\rho) + \tau Z^1_{\sigma,k}(\rho)}{2} \left(\nabla \bm \phi\right)^2 \nonumber\\
    -& \frac{Z_{\sigma,k}(\rho)-Z_{\pi,k}(\rho)+\tau \left(Z^1_{\sigma,k}(\rho)-Z^1_{\pi,k}(\rho)\right)}{4 \rho} \left(\phi_1^2 \left(\nabla \phi_2\right)^2+\phi_2^2 \left(\nabla \phi_1\right)^2\right) \nonumber\\
        +&T_k(\rho) \phi_1 \phi_2 \nabla \phi_1 \nabla \phi_2 
\Bigg\}\;.
\end{align} 
The index E in $\Gamma_{\mathrm{E}}$ stands for `expanded'.
The action $\Gamma_{\mathrm{E}}$ reduces to the $O(2)$-symmetric form $\Gamma_{\mathrm{ISO}}$ if the functions $U^1_k(\rho)$, $Z_{\sigma,k}^1(\rho)$ and $Z_{\pi,k}^1(\rho)$ vanish identically and additionally $T_k(\rho) \equiv \frac{Z_{\sigma,k}(\rho) - Z_{\pi,k}(\rho)}{2 \rho} \equiv Y_k(\rho)$. The smoothness of $\Gamma_{\mathrm{E}}$ imposes the constraints $Z_{\sigma,k}(0)=Z_{\pi,k}(0)$ and $Z^1_{\sigma,k}(0)=Z^1_{\pi,k}(0)$. We introduce  
\beq
 \mathcal{F}^{\mathrm{E}}_k(\rho) = \left\{U_k(\rho), U_k^1(\rho), Z_{\sigma,k}(\rho), Z_{\sigma,k}^1(\rho), Z_{\pi,k}(\rho), Z_{\pi,k}^1(\rho), T_k(\rho) \right\} 
\eeq
as the set of functions parametrizing the flowing effective action in the $\tau$-expanded truncation. For the expanded version (the 2nd approach) we use the variables $(\rho,\tau)$ which allow to easily implement the symmetries and preserve the regularity of the effective action. In the case of the full ansatz (the 1st approach) the use of the variables $(\rho,\tau)$ would give rise to complicated boundary conditions. In view of this we implement the variables $(\rho,\theta)$, for which the boundary conditions are simpler.

The two approximations are complementary and yield very similar results in $d=3$, as we demonstrate below. Differences appear upon varying $d$ towards two. The non-expanded variant of the calculation, requiring a two-dimensional field grid, is clearly more demanding numerically. Both variants of the approximation give the proper critical exponents in $d=4-\epsilon$, up to corrections of order $\epsilon^2$. 

\subsection{Flow equations}
The derivation of the flow equations is a conceptually straightforward generalization of the procedure applied previously for the $O(N)$ models. Implementing the preimposed ansatz, one starts off by evaluating the matrix of second derivatives at a spatially uniform field configuration
\beq 
\Gamma^{(2)}_{X,\alpha_1,\alpha_2}(p^2)=\left.\frac{\delta^2\Gamma_X}{\delta\phi_{\alpha_1}(p)\delta\phi_{\alpha_2}(-p)}\right|_{\text{Uniform}}
\eeq 
 with $X\in\{F, E\}$ and $\alpha_1$, $\alpha_2 \in \{1,2\}$. In contrast to the pure $O(N)$ model, $\Gamma^{(2)}_X$ contains also components off-diagonal in the field index. The obtained result is subsequently supplemented with the cutoff $R_k(q)\delta_{\alpha_1,\alpha_2}$ and inverted. Plugging the resulting formula into the right-hand side of Eq.~(\ref{Wetterich_eq}), one obtains the flow equation for the effective potential $U_k(\rho, \theta)$, which is given in  Eq.~\eqref{LPA}: 
 \begin{widetext} 
\begin{align}
\label{LPA}
%&k \partial_k U_k(\rho, \theta) = \beta^U_k(\rho, \theta) = \int d^dq k \partial_k R_k\left(q\right)  
%\frac{2 R_k(q) + \Gamma^{(2)}_{11}(q^2)+\Gamma^{(2)}_{22}(q^2)}{\left(R_k(q) + \Gamma^{(2)}_{11}(q^2)\right)\left(R_k(q) + \Gamma^{(2)}_{22}(q^2)\right) - \Gamma^{(2)}_{12}(q^2)^2}\;, \nonumber\\
%&\Gamma^{(2)}_{11}(q^2) = q^2\left(Z(\rho ,\theta )+  \cos (2 \theta ) \rho W(\rho ,\theta )\right)
%+U^{(1,0)}(\rho ,\theta )+2 \rho  \cos ^2(\theta ) U^{(2,0)}(\rho ,\theta )
%+\frac{\sin^2 (\theta ) U^{(0,2)}(\rho ,\theta )+ \sin (2\theta ) \left(U^{(0,1)}(\rho ,\theta )-2 \rho  U^{(1,1)}(\rho ,\theta )\right)}{2 \rho } \nonumber\;,\\
%&\Gamma^{(2)}_{22}(q^2) = q^2\left(Z(\rho ,\theta )-  \cos (2 \theta ) \rho W(\rho ,\theta )\right)
%+U^{(1,0)}(\rho ,\theta )+2 \rho  \sin ^2(\theta ) U^{(2,0)}(\rho ,\theta )
%+\frac{\cos ^2(\theta ) U^{(0,2)}(\rho ,\theta )-\sin (2 \theta ) \left(U^{(0,1)}(\rho ,\theta ) - 2 \rho  U^{(1,1)}(\rho ,\theta )\right)}{2 \rho } \nonumber\;,\\
%&\Gamma^{(2)}_{12}(q^2) = q^2 \rho T(\rho,\theta)-\frac{\sin (2\theta ) \left(U^{(0,2)}(\rho ,\theta )-4 \rho ^2 U^{(2,0)}(\rho ,\theta )\right)}{2 \rho }
%+\frac{\cos (2 \theta ) \left(U^{(0,1)}(\rho ,\theta )-2 \rho  U^{(1,1)}(\rho ,\theta )\right)}{2 \rho }\;. 
&k \partial_k U_k(\rho, \theta) = \beta^U_k(\rho, \theta) = \frac{1}{2}\int \frac{d^dq}{(2\pi)^d} k \partial_k R_k\left(q\right)  
\frac{2 R_k(q) + \Gamma^{(2)}_{11}(q^2)+\Gamma^{(2)}_{22}(q^2)}{\left(R_k(q) + \Gamma^{(2)}_{11}(q^2)\right)\left(R_k(q) + \Gamma^{(2)}_{22}(q^2)\right) - \Gamma^{(2)}_{12}\left(q^2\right)^2}\;, \nonumber\\
&\Gamma^{(2)}_{11}(q^2) = q^2\left(Z(\rho ,\theta )+  \cos (2 \theta ) \rho W(\rho ,\theta )\right)
+U^{(1,0)}(\rho ,\theta )+2 \rho  \cos ^2(\theta ) U^{(2,0)}(\rho ,\theta )
+\frac{\sin^2 (\theta ) U^{(0,2)}(\rho ,\theta )+ \sin (2\theta ) \left(U^{(0,1)}(\rho ,\theta )-2 \rho  U^{(1,1)}(\rho ,\theta )\right)}{2 \rho } \nonumber\;,\\
&\Gamma^{(2)}_{22}(q^2) = q^2\left(Z(\rho ,\theta )-  \cos (2 \theta ) \rho W(\rho ,\theta )\right)
+U^{(1,0)}(\rho ,\theta )+2 \rho  \sin ^2(\theta ) U^{(2,0)}(\rho ,\theta )
+\frac{\cos ^2(\theta ) U^{(0,2)}(\rho ,\theta )-\sin (2 \theta ) \left(U^{(0,1)}(\rho ,\theta ) - 2 \rho  U^{(1,1)}(\rho ,\theta )\right)}{2 \rho } \nonumber\;,\\
&\Gamma^{(2)}_{12}(q^2) = q^2 \rho T(\rho,\theta)-\frac{\sin (2\theta ) \left(U^{(0,2)}(\rho ,\theta )-4 \rho ^2 U^{(2,0)}(\rho ,\theta )\right)}{4 \rho }
-\frac{\cos (2 \theta ) \left(U^{(0,1)}(\rho ,\theta )-2 \rho  U^{(1,1)}(\rho ,\theta )\right)}{2 \rho }\;.
\end{align}
\end{widetext} 
For clarity, we suppressed the $k$ dependence in the parametrizing functions and their derivatives.  At leading (zeroth) order of the DE, the local potential approximation [$Z_k(\rho,\theta)=1$, $W_k(\rho,\theta)=T_k(\rho,\theta)=0$], the flow of gradient terms in $\Gamma_X$ is disregarded such that the flow equation for $U_k(\rho, \theta)$ is closed. In the $\Gamma_{\textrm E}$ scheme we additionally perform an expansion of the flow equation for $U_k(\rho, \tau)$ in the $\tau$ direction, while in the $\Gamma_{\textrm F}$ scheme, Eq.~\eqref{LPA} is studied without further approximations. 

In each of the schemes, when going to the level of DE2, the flow of $U_k(\rho, \theta)$ must be supplemented by the flow equations parametrizing the momentum-dependent components of the propagators. These are obtained by first taking the second (functional) derivative of Eq.~(\ref{Wetterich_eq}), which yields the flow of the two-point function $\Gamma^{(2)}_{X, \alpha_1, \alpha_2}(p^2)$. The resulting expressions are subsequently expanded in (external) momentum up to order $p^2$, which projects out the flow equations for the functions $Z_k(\rho, \theta)$, $T_k(\rho, \theta)$, $W_k(\rho,\theta)$ used in the $\Gamma_{\mathrm{F}}$ scheme. Additional expansion in $\tau$ yields the flow equations in the $\Gamma_{\mathrm{E}}$ scheme. 

The derivation procedure is quite mechanical, but the resulting expressions appear very lengthy and rather useless for a human reader. We refrain from quoting them in the text. They are available online at https://doi.org/10.18150/MNMJAO.

\subsection{Fixed points} 
The obtained flow equations are subsequently brought to a scale-invariant form admitting the possibility of fixed-point behavior. This parallelizes the previous treatment of the isotropic $O(N)$ models and is achieved by rescaling the invariants and the parametrizing functions by the running scale raised to the power of the canonical dimension, e.g. $\tilde \rho = \rho k^{-d_\rho}$. The dimensions of the invariants and functions are as follows:
\begin{align}
&d_\rho = d-2+\eta_k\;, \quad d_\tau = 2d -4 +2\eta_k\;, \quad d_\theta = 0\;, \\
&d_U = d\;, \quad d_Z = d_{Z_\sigma} = d_{Z_\pi} = -\eta_k\;,\quad d_W=d_Y = d_T = 2-d-2\eta_k\;,\nonumber \\
&d_{U^1} = 4-d + 2\eta_k\;, \quad d_{Z^1_\sigma} = d_{Z^1_\pi} = 4-2d-3\eta_k\;. \nonumber
\end{align}
Above, $\eta_k$ is the running anomalous dimension defined as a logarithmic derivative of the scaling factor $\eta_k = -k \partial_k \log( Z_k)$. To close the set of equations we define the scaling factor $Z_k$ by imposing the condition  $Z(\rho_\eta)/Z_k = \tilde Z(\tilde\rho_\eta) \equiv 1$ for an arbitrary constant $\tilde \rho_\eta$. It is also convenient to introduce the `renormalization time' $t=\log(k/\Lambda)$ [$\partial_t = k\partial_k$], which allows to express the flow equations in a form invariant upon RG time translations. After rescaling, Eq.~\eqref{LPA} takes form:
\begin{equation}
\partial_t \tilde{U}_t(\tilde \rho, \theta) = -d \tilde{U}_t(\tilde \rho, \theta) + (d-2+\eta_k) \tilde \rho \tilde U'_t(\tilde \rho, \theta)+\tilde{\beta}^U_t(\tilde \rho, \theta),
\end{equation}
with $\tilde \beta^U_t(\tilde \rho, \theta) = k^{-d_U} \beta^U_k(\rho,\theta)$.

In the numerical treatment, the parametrizing functions are represented by their values on a finite grid of points in the field space. The grid is two-dimensional within the $\Gamma_{\mathrm{F}}$ scheme, and one-dimensional within the $\Gamma_{\mathrm{E}}$ approximation. In the $\theta$ direction, the grid contains 21 points. The size of the grid in the $\tilde \rho$ direction depends on the dimension, ranging from 40 points in $d=3$ to 120 in $d=2$, as the structure of the fixed point becomes more complex the lower the dimension. To ensure cubic symmetry of $\Gamma_{\mathrm F}$, we have to impose $\mathcal{F}_k^{\mathrm{F}}(\rho,\theta)$ to be a periodic function of $\theta$ with period $\pi/2$ and even upon reflection with respect to $\pi/4$. We implement this by considering the $\theta$-grid on an interval $[0,\pi/4]$. The periodicity and reflection-symmetry conditions additionally allow us to approximate derivatives with central finite differences instead of forward (backward) finite differences even at the boundary of the $\theta$ grid.

   The derivatives occurring in the equations are approximated with finite differences and integrals with finite sums. In such a way we cast the set of integro-differential equations onto a huge set of algebraic  equations which can be solved numerically. 

The presently studied universal quantities are extracted from the structure of the flow equations around the fixed point. First, we find the fixed point by solving the equation:
\begin{equation}
    \partial_t \mathcal{F}_{\mathrm{ISO}}^* =0\;,
\end{equation}
using the Newton's method. In the present paper, we are only interested in the structure of the flow in the vicinity of the $O(2)$-symmetric fixed points. Therefore, we only need to solve a  significantly simpler equation for $\Gamma_{\mathrm{ISO}}$. This alone is enough to find the anomalous dimension $\eta$. For the other exponents we introduce the stability matrix:
\begin{equation}
    \mathcal{M}_{i,\tilde \rho; j,\tilde \rho'} := \left.\frac{\partial\left(\partial_t \tilde{\mathcal{F}}_i(\tilde{\rho})\right)}{\partial \tilde{\mathcal{F}}_j(\tilde{\rho}')}\right|_{\mathcal{F}^*},
\end{equation}
which represents the linear expansion of the flow equations around the fixed point $\mathcal{F}^*$. The symbol $\tilde{\mathcal{F}}_i(\tilde{\rho})$ denotes the value of the $i$-th function in the vector $\tilde{\mathcal{F}}$ evaluated at the point~$\tilde{\rho}$ [or $(\tilde{\rho}, {\theta})$ on 2-dimensional grid]. From this matrix we remove the rows and columns corresponding to values fixed by external constraints, e.g. we remove row and column corresponding to $\tilde Z(\tilde \rho_\eta)$ because its value is fixed to $1$ to extract the anomalous dimension. 

The eigenvalues of the stability matrix are related to the critical exponents; $e_1 = \frac{1}{\nu}$ - the leading (and only positive) eigenvalue is the inverse of the correlation length exponent, while the lower eigenvalues define the correction to scaling exponents $\omega_i = |e_{i+1}|$ \cite{Pelissetto_2002}. The stability matrix is not symmetric and therefore its eigenvalues may in general be complex. The complex eigenvalues always appear as conjugate pairs: $\lambda$ and $\bar{\lambda}$.

%In what follows, we implement the procedure described above within the two introduced NPRG truncations to provide the estimates of the eigenvalue $y_4$ in dimensionality $d=3$, together with the error bar (Sec.~IIIA). 
Throughout the analysis we implement two cutoff functions:
\begin{alignat}{2}
&\text{Wetterich}\quad &&R^{\mathrm{W}}_k(q)=\frac{\alpha Z_k q^2}{e^{q^2/k^2}-1} \\
&\text{Exponential}\quad &&R_k^{\mathrm{E}}(q)=\alpha Z_k k^2 e^{-q^2/k^2} \nonumber
\label{Wetterich_cutoff}
\end{alignat} 
with a variable parameter $\alpha$. In accord with the principle of minimal sensitivity \cite{Canet_2003_2, Balog_2020} (PMS) we associate the physical values of the computed quantities with values obtained at $\alpha$ such that the quantity in question is locally stationary with respect to variation of $\alpha$. We verified that, once the values of $\alpha$ are fixed by implementing the PMS procedure, both cutoffs yield almost identical results. For that reason, all the presented figures only include data obtained with the Wetterich cutoff.

%Having treated the case $d=3$ in Sec.~IIIA, subsequently, in Sec.~IIIB we will gradually reduce $d$ in order to investigate the interplay between the leading irrelevant eigenvalues (both related to $O(2)$-symmetric and to $\mathbb{Z}_4$-symmetric perturbations) with particular focus on the limit $d\to 2^+$.

\section{Results} 
The neighborhood of $d=4$ is well described by perturbative means. As already mentioned, in this regime, the approximations implemented by us are exact up to corrections of order $\epsilon^2$. We therefore focus here on the range $d\in [2,3]$. 

We implement the procedure described above in Sec.~II, treating $d=3$ and $d<3$ separately. In the former case (Sec.~IIIA), we aim at providing numerically accurate estimates of the leading irrelevant eigenvalues, applying the error estimate methodology developed in Refs.~\onlinecite{Polsi_2020, Polsi_2021} and summarized in Sec.~IIIC. For $d<3$ (Sec.~IIIB) we aim at capturing the approach of the subleading eigenvalues towards zero and the emergence of the KT physics as well as nonuniversal critical behavior at $h_4\neq 0$. We compare the two truncations described above (the $\Gamma_{\textrm{F}}$ and $\Gamma_{\textrm{E}}$ schemes) and point out that while in $d=3$ their predictions are practically equivalent, important differences appear at lower $d$. 
\subsection{Dimensionality $d=3$} 
\label{resultsd3}
In Fig.~1 we plot the leading stability matrix eigenvalues obtained at order DE2 within the $\Gamma_{\textrm{F}}$ and $\Gamma_{\textrm{E}}$ schemes as function of the cutoff parameter $\alpha$. The former are exhibited with lines, the latter with points. The two sets of data practically coincide and their variation as function of $\alpha$ is hardly visible in the plot scale. The eigenvalues may be divided into those related to the isotropic component of the theory [i.e. existent also in the isotropic case - Eq.~(\ref{Ansatz_iso})] and those appearing only in the $\mathbb{Z}_4$-symmetric setup. Members of the former category are supplied with the superscript `iso', the latter with `aniso'.
\begin{figure}[ht]
\begin{center}
\includegraphics[width=9.0cm]{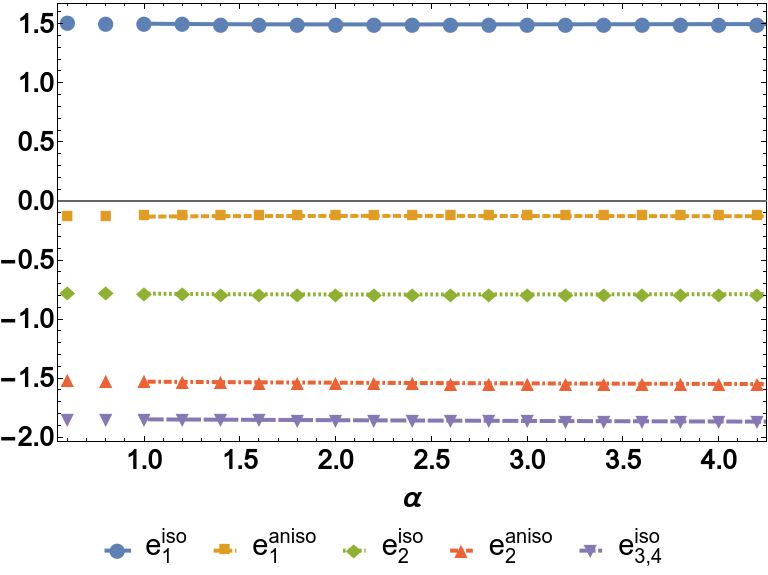}
\caption{(Color online)  The leading stability matrix eigenvalues obtained at order DE2  displayed as function of the cutoff parameter $\alpha$ in $d=3$. The results obtained within the $\Gamma_{\textrm{F}}$ and $\Gamma_{\textrm{E}}$ schemes are exhibited with lines, and points respectively. The relevant eigenvalue $e_1^{\textrm{iso}}$ determines the correlation length exponent $\nu$. The leading irrelevant eigenvalue $y_4=e_1^{\textrm{aniso}}$ emerges due to the anisotropy; the dominant irrelevant isotropic eigenvalue $e_2^{\textrm{iso}}$ is significantly further from zero as compared to $e_1^{\textrm{aniso}}$. The results obtained within the two schemes practically coincide.}
\label{eigenvals_3d}
\end{center}
\end{figure} 
The eigenvectors corresponding to the eigenvalues labeled as `iso' preserve the $O(2)$ symmetry, those labeled with 'anizo' do not.  In accord with the expectations, the leading irrelevant exponent $y_4=e_1^{\textrm{aniso}}$ arises due to anisotropy and is situated significantly closer to zero as compared to the leading irrelevant isotropic eigenvalue $e_2^{\textrm{iso}}$. The quantities $e_1^{\textrm{iso}}$ and $e_2^{\textrm{iso}}$ determine the correlation length exponent $\nu$ as well as the correction to scaling exponent $\omega$ of the isotropic $O(2)$ model. The present values of these quantities coincide with those obtained previously at the level of DE2 in Refs.~\onlinecite{Polsi_2020, Chlebicki_2021}. While the first few leading eigenvalues are real,  some of the more irrelevant eigenvalues obtained by us contain an imaginary component. In all figures of the present paper, we plot only the real part of eigenvalues; the pairs of complex eigenvalues are distinguished by a double subscript of the data series label, e.g. $e^{\textrm{iso}}_{3,4}$. Among the quantities exhibited in Fig.~\ref{eigenvals_3d} all are real except $e_{3,4}^{\textrm{iso}}$, which contain a small imaginary component. 
%Only the real parts of the eigenvalues are displayed in the plots.

\begin{figure}[ht]
\begin{center}
\includegraphics[width=9cm]{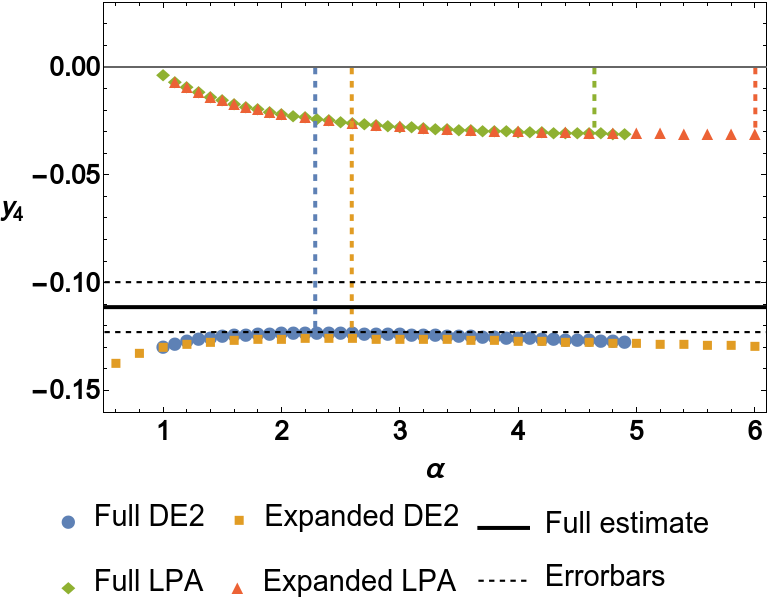}
\caption{(Color online) The dependence of $y_4$  on the cutoff parameter $\alpha$, comparing the LPA and DE2 approximation levels in the full and $\tau$-expanded schemes in $d=3$. The vertical dashed lines indicate the PMS values corresponding to the four sets of data. The bold line indicates our final estimate of $y_4=-0.111$ at the level of DE2 and is accompanied by the error bars (horizontal dashed lines).  }
\label{y4_3d}
\end{center}
\end{figure}

In Fig.~\ref{y4_3d} we plot the dominant irrelevant eigenvalue $y_4=e_1^{\textrm{aniso}}$ obtained at zeroth (LPA) and first (DE2) approximation levels within the $\Gamma_{\textrm{E}}$ and $\Gamma_{\textrm{F}}$ schemes. There is practically no difference between the results found from these two schemes. 
Note that at LPA level $y_4$ changes sign as a function of $\alpha$, while its PMS value is (in its modulus) underestimated by a factor of order 4 with respect to the anticipated range of values. In contrast, the values obtained at order DE2 remain well separated from zero for all values of $\alpha$. The final estimate of $y_4$ at order DE2 (together with the error bars) is obtained by means of the procedure summarized in Sec.~IIIC. We postpone the discussion of these methodological aspects to Sec.~IIIC. We note here that our present error estimate methodology is very conservative. In Table I we present a comparison between our result and the values obtained from a diversity of approaches including perturbation theory (at six loops), MC simulations, large charge expansion, as well as earlier NPRG results. We observe a substantial spread of the reported values,  considering the available error estimates.   
\begin{table}[h!]
\centering
\begin{tabular}{ |c|c| } 
 \hline
 PT (6 loop) (Ref.~\onlinecite{Carmona_2000}) &-0.103(8) \\ 
 MC (Refs.~\onlinecite{Hasenbusch_2011, Okubo_2015}) & -0.108(6) \\ 
 LCE (Ref.~\onlinecite{Banerjee_2018}) & -0.128(6) \\ 
 MC/RG (Ref.~\onlinecite{Shao_2020}) & -0.114(2) \\
 LPA' (Ref~\onlinecite{Leonard_2015}) & -0.042  \\  
 DE2 (This work) & -0.111(12) \\
 \hline 
\end{tabular} 
 \caption{%{\it Table partially stolen from Ref~.\onlinecite{Shao_2020}. Check original papers, look for more. [CFT and MC inconsistent ?]} 
 Comparison of the values of $y_4$ obtained within different theoretical and simulation approaches including perturbation theory (PT), Monte Carlo (MC) simulations and large charge expansion (LCE). Note a substantial spread of the values, in particular the differences between the LCE and MC / PT predictions.  } 
 \label{Table_1} 
 \end{table}
 
We finally point out that an accurate parametrization of the momentum dependencies is crucial for an accurate computation of $y_4$ within the NPRG methodology. While our results obtained at order DE2 are of accuracy comparable to the other approaches listed in Table 1, one may expect that (if performed) a calculation reaching the order DE4 would deliver numbers even more accurate.   

\subsection{Dimensionality $d<3$} 
\label{resultslowd}
The NPRG methodology allows for treating the dimensionality $d$ as a continuously variable parameter. In the present study it creates the possibility of smoothly interpolating between $d=3$ and $d=2$, two situations involving completely different physics of critical phenomena. We therefore gradually depart from $d=3$ and follow the evolution of the stability matrix eigenvalues as $d$ is decreased. Our results concerning the DE2 level PMS values of the stability matrix eigenvalues as obtained within the two NPRG truncations are summarized in Fig.~3.  In both schemes we succeeded in covering the entire range $d\in [2,3]$. Even though we are able to find fixed points in $d=2$, the PMS criterion for fixing the dependence on $\alpha$ of the various critical exponents can not be applied in that dimension. In that dimension, the criterion for fixing $\alpha$ proposed in the literature is different \cite{Jakubczyk_2014} due to the physical existence of a line of fixed points. For this reason we present the results for dimensions strictly larger than $d=2$, approaching this value from above. As discussed in Sec.~IIIA, in the vicinity of $d=3$ our results obtained within the two truncation schemes may be treated as practically identical. Significant differences appear for $d\lesssim 2.5$, where crossing of eigenvalues occurs within the $\Gamma_E$ approach, but not in $\Gamma_F$. 

As concerns the relevant eigenvalue $e_1^{\textrm{iso}}$, we recover the monotonous approach towards zero as $d\to 2^+$, indicated the onset of the KT physics and the essential singularity of the correlation length \cite{Chaikin_book}. Interestingly, the leading irrelevant eigenvalue $e_1^{\textrm{aniso}}=y_4$ initially increases in its absolute value as $d$ is reduced below $d=3$. This happens within both the calculations performed by us and is also robust with respect to cutoff variation. Such behavior indicates (since $y_4$ vanishes in $d=2$) that the dependence of the leading anisotropic eigenvalue, $y_4$ on $d$ cannot be monotonous. The scenario is slightly different in the two schemes. Within the $\Gamma_{\textrm{E}}$ approach (see Fig.~3.b) $y_4=e_1^{\textrm{aniso}}$ exhibits a minimum at $d\approx 2.2$ and approaches zero upon reducing $d=2$, which indicates the appearance of symmetry-breaking phase transitions characterized by non-universal critical exponents \cite{Jose_1977}. Note the crossing of eigenvalues which occurs in the $\Gamma_{\textrm{E}}$ approach but not in the $\Gamma_{\textrm{F}}$ one. We additionally note that, within the $\Gamma_{\textrm{E}}$ approach,  the eigenvalues $e_2^{\textrm{aniso}}$ and $e_3^{\textrm{aniso}}$ acquire an imaginary component below $d\approx 2.5$ and are mutually complex conjugate below $d\approx2.5$. For $d\to 2^+$ they also become very close to zero. In the $\Gamma_{\textrm{F}}$ approach (see Fig.~3.a) the anisotropic eigenvalue $e_1^{\textrm{aniso}}$ has a very sharp drop below $d\sim 2.03$ which points towards reaching a value near zero in $d=2$ in a very strongly singular way. Indeed, in the lowest dimension where we were able to apply the PMS procedure, $d=2.01$, we reach the relatively large value $-0.25$ but the value of $e_1^{\textrm{aniso}}$ at $d=2.03$ is as large as $-0.4$\footnote{We actually verified in $d=2$ that $y_4$ goes to zero in a value of $\alpha$ where $\eta \cong 1/4$.}.
 It must be said, in any case that given the large uncertainties in those dimensions (see next Section), we can safely say that the obtained eignenvalue is compatible with zero. 
\begin{widetext} 

\begin{figure}
%\begin{center} 
\begin{subfigure}[t]{8.5cm}
\centering\includegraphics[width=8.4cm]{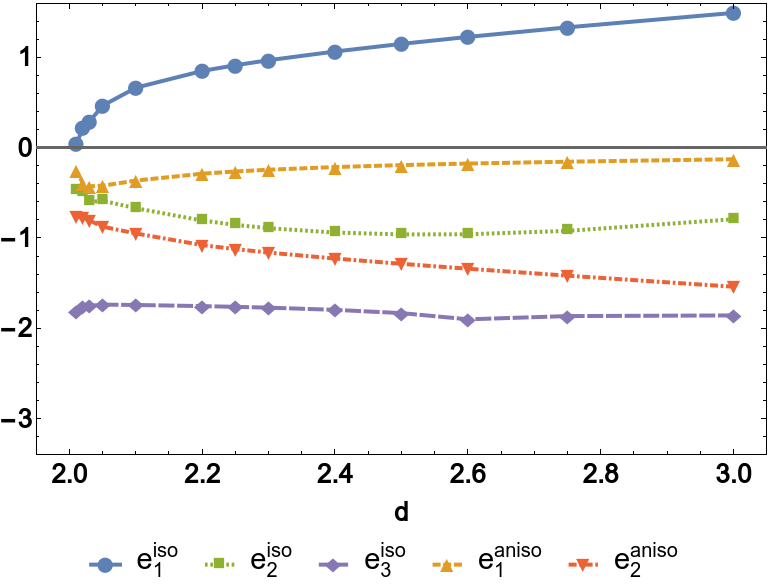}
\caption{(Color online) Evolution of the PMS results for the leading stability matrix eigenvalues upon varying $d$ as obtained within the $\Gamma_{\textrm{F}}$ scheme. Down to $d\approx 2.5$ the eigenvalues obtained within the $\Gamma_{\textrm{F}}$ scheme are very close to those resulting from the $\Gamma_{\textrm{E}}$ scheme (compare Fig.~4). Differences arise in the range $d\in [2.0, 2.5]$. In contrast to the  $\Gamma_{\textrm{E}}$ case, there are no eigenvalues' crossings. For $d\to 2^+$ the eigenvalues $e_1^{\textrm{aniso}}$
and $e_1^{\textrm{iso}}$ approach zero in a singular fashion.}
%\label{PMS_full}
%\end{center}
\end{subfigure} 
\begin{subfigure}[t]{8.5cm}
%\begin{center}
\centering\includegraphics[width=8.4cm]{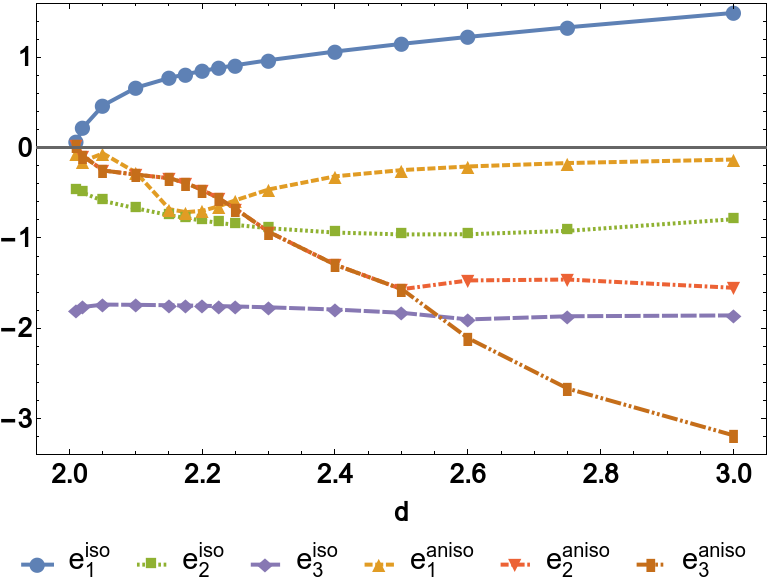}
\caption{(Color online) Evolution of the PMS results for the leading stability matrix eigenvalues upon varying $d$ as obtained within the $\Gamma_{\textrm{E}}$ scheme. The relevant eigenvalue $e_1^{\textrm{iso}}$ monotonously approaches zero for $d\to 2^+$, marking the onset of the KT physics characterized by the essential singularity of the correlation length. The leading anisotropic eigenvalue $e_1^{\textrm{aniso}}$ becomes very close to zero as $d\to 2^+$ indicating the appearance of symmetry-breaking phase transitions characterized by non-universal critical exponents \cite{Jose_1977}. Its dependence on $d$ is non-monotonous. The eigenvalues $e_2^{\textrm{aniso}}$ and $e_3^{\textrm{aniso}}$ acquire an imaginary component below $d\approx 2.5$ and are mutual complex conjugate below $d\approx2.5$. For $d\to 2^+$ they become very close to zero. Note the crossings between the eigenvalues, absent in the $\Gamma_{\textrm{F}}$ calculation and presumably appearing as a truncation artefact.}
%\label{PMS_expanded}
%\end{center}
\end{subfigure} 
\caption{(Color online)  Comparison between the leading PMS stability matrix eigenvalues obtained within the $\Gamma_{\textrm{F}}$ and $\Gamma_{\textrm{E}}$ schemes, varying dimensionality between $d=3$ and $d=2$.    }
\end{figure} 
 
\end{widetext}
The accuracy control of our approach is much better in $d=3$ than in the vicinity of $d=2$ (see Sec. IV for further discussion). In the present calculation this manifests itself with the differences appearing in the results obtained within the two truncation schemes. We point out however, that the present NPRG implementation captures the subtle effects arising in $d=2$ due to the appearance of marginal operators, and the role of vortices, controlling the KT transition. This, together with the accurate resolution of $y_4$ in the vicinity of $d=3$ constitutes a significant improvement as compared to the previous NPRG studies of this system. We observe that the behavior of $y_4$ as function of $d$ is nonmonotonous and exhibits a minimum slightly above $d=2$ (the precise turning point depending on the truncation). Both in the $\Gamma_E$ and the $\Gamma_F$ schemes $y_4$ exhibits an apparent approach towards zero for $d\to 2$.

In Fig.~4 we demonstrate the dependence of the leading anisotropic eigenvalues on $\alpha$, comparing the results from the $\Gamma_F$ and $\Gamma_E$ schemes for a sequence of values of $d$. While the two data sets practically coincide in the vicinity of $d=3$, quantitative differences occur below $d\approx 2.5$, while for $d\approx 2.2$ the entire picture is substantially different due to the eigenvalues' crossing occurring in the $\Gamma_E$ scheme. In Fig.~4(e) we present the plot corresponding to $d=2.01$, since for $d=2.0$ there is no clear PMS value and a different optimization procedure needs to be implemented \cite{Jakubczyk_2014}.  
\begin{widetext} 

\begin{figure}
  \begin{subfigure}{8.5cm}
    \centering\includegraphics[width=8.4cm]{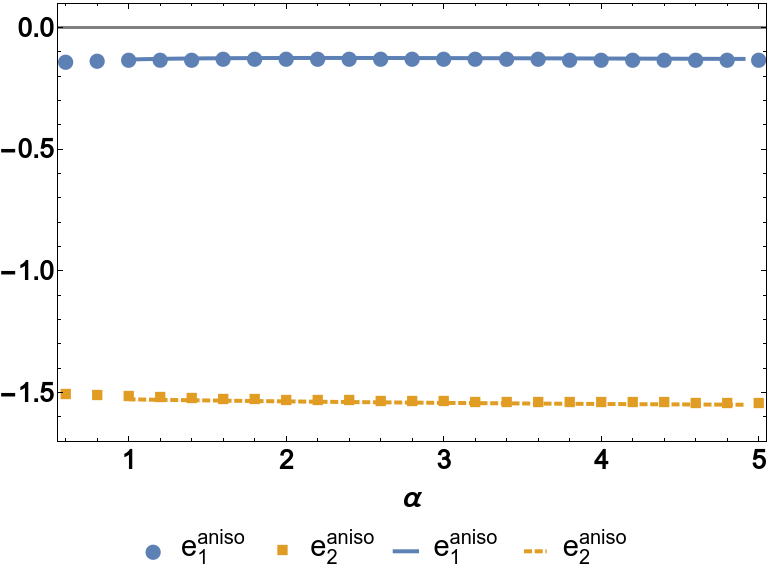}
%    \caption{$d=3$}
\subcaption{$d=3$.}
  \end{subfigure}
  \begin{subfigure}{8.5cm}
    \centering\includegraphics[width=8.4cm]{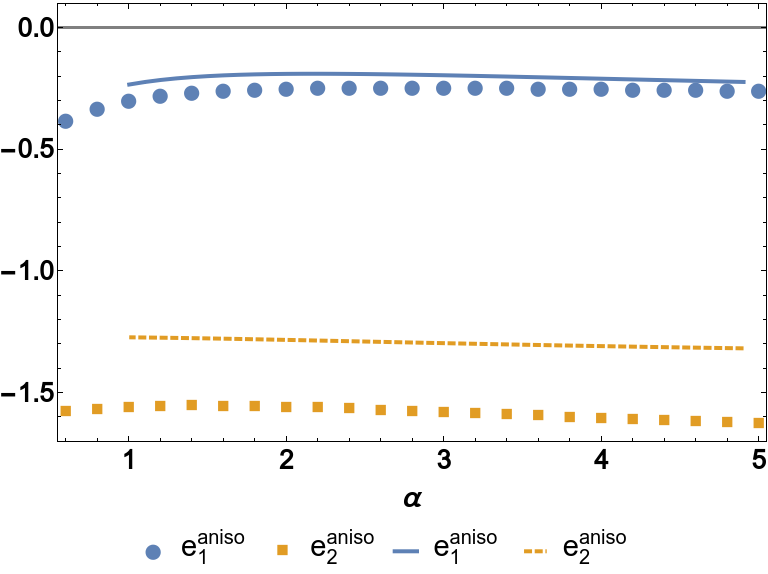}
%    \caption{$d=2.5$}
\subcaption{$d=2.5$.}
  \end{subfigure}
  
\vspace{.3cm}
  \begin{subfigure}{8.5cm}
    \centering\includegraphics[width=8.4cm]{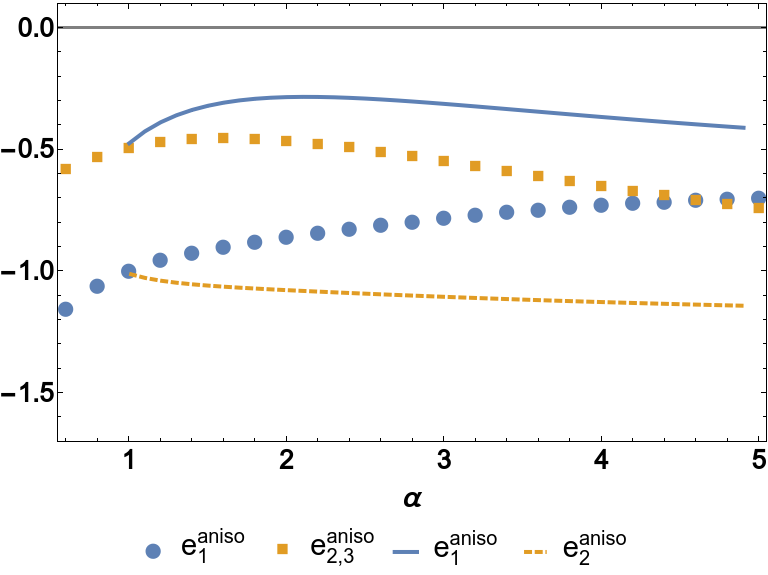}
%    \caption{$d=2.2$}
\subcaption{$d=2.2$.}
  \end{subfigure}
  \begin{subfigure}{8.5cm}
    \centering\includegraphics[width=8.4cm]{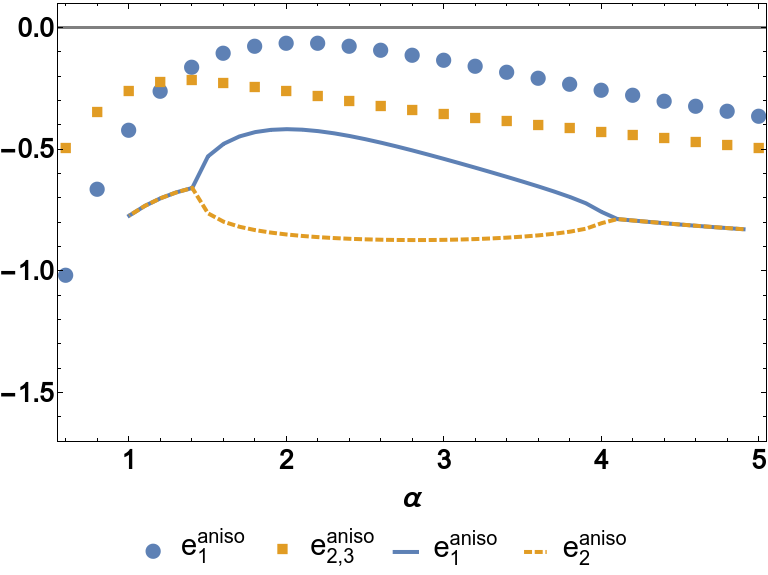}
\subcaption{$d=2.05$.}    
 %   \caption{$2.05$}
  \end{subfigure}
  
\vspace{.3cm}
  \begin{subfigure}{8.5cm}
    \centering\includegraphics[width=8.4cm]{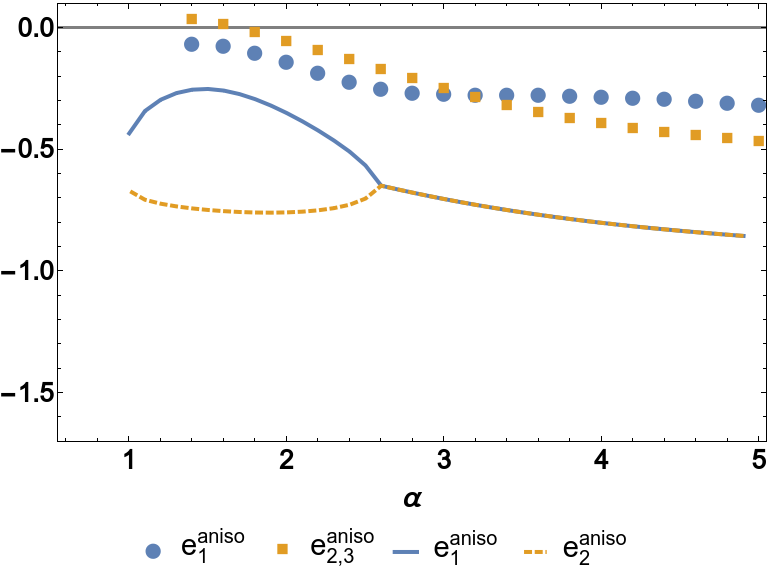}
\subcaption{$d=2.01$.}    
 %   \caption{$2.05$}
  \end{subfigure}   
  \caption{(Color online)  The leading stability matrix eigenvalues related to the anisotropy obtained at order DE2  displayed as function of the cutoff parameter $\alpha$ for a sequence of  values of $d$. The results obtained within the $\Gamma_{\textrm{F}}$ and $\Gamma_{\textrm{E}}$ schemes are exhibited with lines, and points respectively. The values obtained within the two schemes practically coincide in $d=3$, but the differences increase when lowering $d$. For $d=2.2$ the leading irrelevant eigenvalue $y_4$ corresponds to $e_1^{\textrm{aniso}}$ within the $\Gamma_{\textrm{F}}$ scheme, and to $e_{2,3}^{\textrm{aniso}}$ within the $\Gamma_{\textrm{E}}$ scheme [compare also Fig.~3(b)]. For $d$ approaching 2, the PMS value of $y_4= e_1^{\textrm{aniso}}$ becomes very close to zero. In figures (d) and (e) some eigenvalues are real and different near the PMS but cross and become complex conjugates far from the PMS. This results in some bifurcations in the curves in those figures.}
\end{figure}
\end{widetext}
\subsection{Error estimates methodology} 
\label{methods}
The procedure that we employed to calculate DE error bars was proposed and explained in detail previously in Ref.~\onlinecite{Polsi_2020}. It has been successfully tested for the calculation of critical exponents in that reference and for universal amplitude ratios in Ref.~\onlinecite{Polsi_2021} (in both cases for $O(N)$ models in $d=3$). We refer to those references for more details and limit ourselves here, in order to make the reading minimally self-contained, to reviewing the main ideas.

The key element for estimating error bars in the DE framework is that successive orders of this approximation are suppressed\footnote{Strictly speaking this observation only take place once one choose as estimate at a given order of the DE the value obtained from the PMS procedure.} by a small parameter of the order of $1/9$--$1/4$ \cite{Balog_2019}. Due to this fact, if we call $Q^{(s)}$ the raw estimate of a certain quantity $Q$ at order $s$ of the DE, a pessimistic generic estimate of the error made within the DE at that order is
\beq
\Delta^{gen} Q^{(s)}=\frac{|Q^{(s)}-Q^{(s-2)}|}{4}.
\eeq
Put another way, we can estimate the standard error at a given order by taking the difference between the estimate at that order with the precendent  and dividing by 4. This is a conservative estimate of the reduction of the uncertainty of successive orders of the DE. It is worth mentioning that this procedure only provides an estimate of the error bar for second and highers orders of the DE.

In many cases, however, this estimate is excessively pessimistic. In particular, it has been observed that  certain critical exponents show in successive orders of the DE an oscillatory behavior that progressively approaches the best estimates in the literature. In such cases, the raw estimate at a given order of the DE should be seen as a bound (upper or lower depending on the case) to the critical exponent under analysis. In this understanding, the values of the previously stated confidence interval that do not satisfy the bound should be discarded and the corresponding confidence interval is divided by a factor of two (and consequently also the central value is shifted slightly in the direction of the previous order of the DE). That is
\begin{align}
\Delta^{impr} Q^{(s)}&=\Delta^{gen} Q^{(s)}/2,\\
Q_{impr}^{(s)}&= Q^{(s)}\pm \Delta^{impr} Q^{(s)}.
\end{align}
In the previous formula the sign is set by shifting the central value towards the estimate of the previous order.
This procedure was also successfully tested in Ref.~\onlinecite{Polsi_2020} giving always precise and accurate results.

In the present work we analyze in detail the behavior of the critical exponent $y_4$ which, as in many other cases, seems to show an alternating behavior and in which the estimate of each order of the DE seems to be a bound (upper in the case of the LPA and lower to the order DE2). Assuming the alternating behavior, we adopt the improved estimation procedure mentioned in the previous paragraph and represented graphically in Fig.~\ref{y4_3d}.

As discussed in Sec.~\ref{resultsd3} the results obtained in this way for $y_4$ in $d=3$ turn out to be in very good agreement with estimates from other methods. Although the precision achieved by us is slightly lower than the best estimates in the literature, it should be noted that at least some of the errors reported in the literature must be underestimated since, as previously pointed out, the reported results are incompatible with each other. It should be noted that the improved procedure for estimating the central value and the confidence interval gives a result that seems to improve the central value. In fact, it is likely that our estimation of error bars is slightly pessimistic but, following the criterion adopted in Ref.~\onlinecite{Polsi_2020} we have preferred to limit ourselves to conservative error bars.

When lowering the dimension by going to $d<3$ the estimated error bars tend to increase significantly. Finally, when reaching the vicinity of $d=2$ the current procedure does not allow us to give a controlled error bar estimate since the dominant order of the DE (the LPA) does not exhibit a fixed point and therefore we only have the central value estimate coming from the DE2 order. It should be noted however that, as pointed out in Sec.~\ref{resultslowd}, the values for various critical exponents obtained in one of the DE2 versions (see below) show a qualitative behavior very similar to the expected one \cite{Jose_1977}.

\section{Ansatz vs strict DE} 
In this section we address a point that was left aside in the methodological discussion of the Sec.~\ref{methods}. In the literature there are two different ways of implementing the DE. The resulting equations are not the same and therefore it is necessary to analyze to what extent the results they give are equivalent. In this section we address this point to order DE2 in the $XY$ model for all dimensions from $d=3$ to $d=2$ for both isotropic exponents and for $\mathbb{Z}_4$-symmetric perturbations.

The most commonly implemented version of the DE which we will refer to as ``ansatz version'' corresponds to simply extracting all vertices and propagators directly from the ansatz of the DE to a given order. For example, in the present work, at order DE2, we simply differentiate Eqs.~(\ref{Ansatz_iso}) or (\ref{double_functional}) an appropriate number of times to obtain propagators and vertices with 3 and 4 legs (the only ones needed to this order of the DE).

A slightly different version of the DE was recently proposed in Ref.~\onlinecite{Balog_2019}. According to the logic developed in that reference it can be expected that all terms at a given order of momenta expansion in a given flow equation have the same order of magnitude. There could be exceptions to this generic estimate but these should have a complementary origin (such as the existence of a symmetry or some limit in which the equations are simplified, e.g., by the existence of a critical dimension). In this understanding, when studying a given flow equation, terms with a number of momenta equal to other terms already neglected by the mere imposition of the ansatz can be neglected. For example, in the study of the two-point function there appear terms that include vertices with 4 legs (tadpole contribution) added to terms involving the product of two vertices with 3 legs (bubble contribution). If one works at order DE2, the terms with 4 momenta are discarded by the ansatz in terms that involve a 4-legs vertex. In consequence, the terms with four momenta appearing in the product of two vertices with three legs can be consistently neglected as well. This is the actual version of the DE that was implemented at high order of the DE in Refs.~\onlinecite{Balog_2019,Polsi_2020,Polsi_2021} and we denote it here as the ``strict'' version of the DE.

Both versions of the DE have their advantages and disadvantages. From the point of view of aesthetics and simplicity of presentation it is clear that the ansatz version is more convenient. Moreover, coming from an ansatz that is taken in the same way for all vertices, all relations coming from symmetries or relations between response functions and correlation functions, are automatically satisfied. On the other hand, the resulting flow equations are more complicated. The difference in the computational complexity is very small at order DE2 for the $O(N)$ models and moderate when including $\mathbb{Z}_4$ perturbations. However, when analyzing the order DE4 (and even more at order DE6) the difference in the size of equations becomes huge and in those cases the simplification of the expressions coming from the ``strict'' version becomes an invaluable help.

The existence of these two versions of the DE generates, however, a difficulty because we must be sure that when implementing one or the other the results are equivalent (within the corresponding error bars). The comparison between these two versions of the DE was previously performed in $d=3$ to order DE2 for the $O(N)$ models and to order DE4 for the $N=1$ case \cite{Polsi_2020}. In Ref.~\onlinecite{Peli_2021} a field-expanded version of the ansatz version is employed at order DE4 for the $O(N)$ models and offers results compatible with those previously obtained in the strict version without field expansion.  In all these cases it is observed that the difference between the two versions of the DE is considerably smaller than the error bars estimated to a given order. These comparisons allow us to conclude that, at least for the purpose of studying universal properties at $d=3$, the two procedures are essentially equivalent.

In the present section we extend these previous studies concerning the comparison between the ansatz and the strict versions of the DE. For this purpose, we compare both versions for the critical $\mathbb{Z}_4$ model at DE2 order and for the first time we perform the comparison not only for dimension 3 but also lower.

The observation to be strongly emphasized is that both versions of the DE are fully compatible at $d=3$ for all the analyzed quantities ($\nu$, $\eta$, $y_4$ in particular). Note that throughout the text we have reported the results coming from the ansatz version (except in the present section when explicitly stated). It is therefore worth discussing here how our estimate of $y_4$ depends on the implementation of the DE. In table~\ref{Table_2} we carry out this comparison analyzing, in addition, the dependency on the family of regulators used, as well as on having used a truncated version in the fields (see Eq.~(\ref{expanded})) or with a complete dependency on $\rho$ and $\theta$ (see Eq.~(\ref{double_functional})). It is clearly observed that in $d=3$ the difference between the strict or ansatz version of the DE is well below the margin of error. As stated in Ref.~\onlinecite{Polsi_2020} a similar behavior is observed at order DE2 for the isotropic exponents $\eta$ and $\nu$. Moreover, in that dimension, the truncation in the anisotropic invariant (when order indicated in the equation Eq.~(\ref{expanded})) or the choice of the regulator family
\footnote{One must, however, stress that the dependence on the regulator is higher without imposing the PMS criterion.} also have effects below the margin of error.

\begin{table}[h!]
\centering
\begin{tabular}{ |c|c|c|c| } 
\hline
Version & Field truncation & Regulator & $y_4$ \\
 \hline
Ansatz & Full & Wetterich & -0.111$\pm$0.012\\
Ansatz & Full & Exponential & -0.112$\pm$0.012 \\
Strict & Full & Wetterich & -0.113$\pm$0.012 \\
Strict & Full & Exponential & -0.114 $\pm$0.012\\
Ansatz & Expanded & Wetterich & -0.114$\pm$0.012\\
Strict & Expanded & Wetterich & -0.114$\pm$0.012\\
\hline 
\end{tabular} 
 \caption{Comparison of the values of $y_4$ obtained within different versions of the DE2 in $d=3$. The field truncation refers to full field content in $\rho$ and $\theta$ (see Eq.~\ref{double_functional}) or a partial expansion by keeping the full dependence in $\rho$ but expanding in the invariant $\tau$ (see Eq.~(\ref{expanded})).} 
 \label{Table_2} 
 \end{table}
 
\begin{figure}[ht]
\begin{center}
\includegraphics[width=8cm]{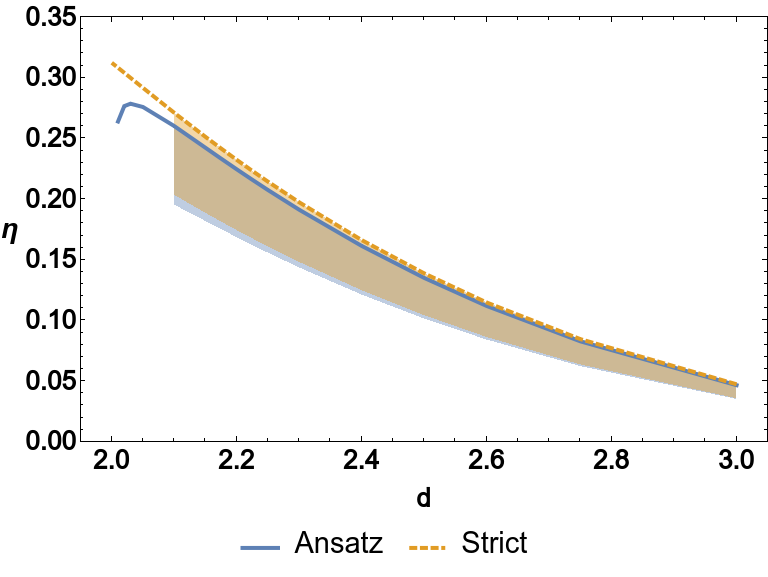}
\caption{(Color online) The dependence of exponent $\eta$ on the implementation of the DE as a function of $d$. Lines represent raw values from the ansatz and strict versions of the DE2. Marked regions represents DE2 confidence intervals for each implementation.}
\label{etaDEversions}
\end{center}
\end{figure}

\begin{figure}[ht]
\begin{center}
\includegraphics[width=8cm]{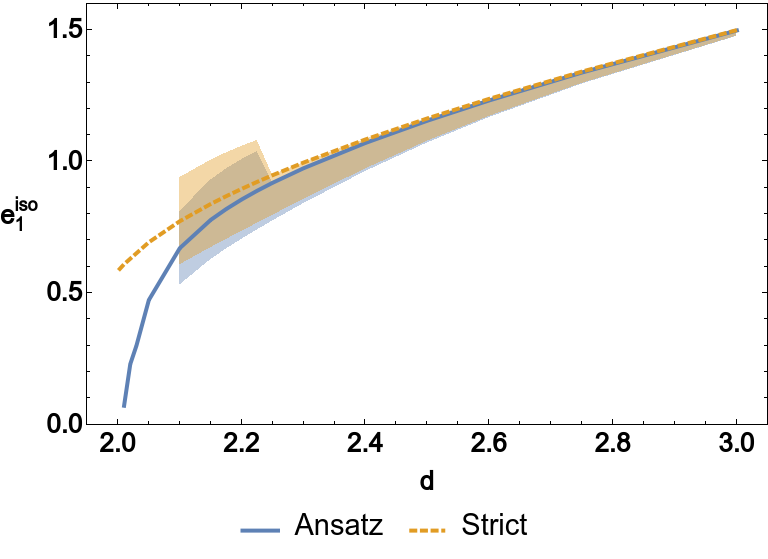}
\caption{(Color online) The dependence of exponent $e_1^{iso}=1/\nu$ on the implementation of the DE as a function of $d$. Lines represent raw values from the ansatz and strict versions of the DE2. Marked regions represents DE2 confidence intervals for each implementation.}
\label{nuDEversions}
\end{center}
\end{figure}

\begin{figure}[ht]
\begin{center}
\includegraphics[width=8cm]{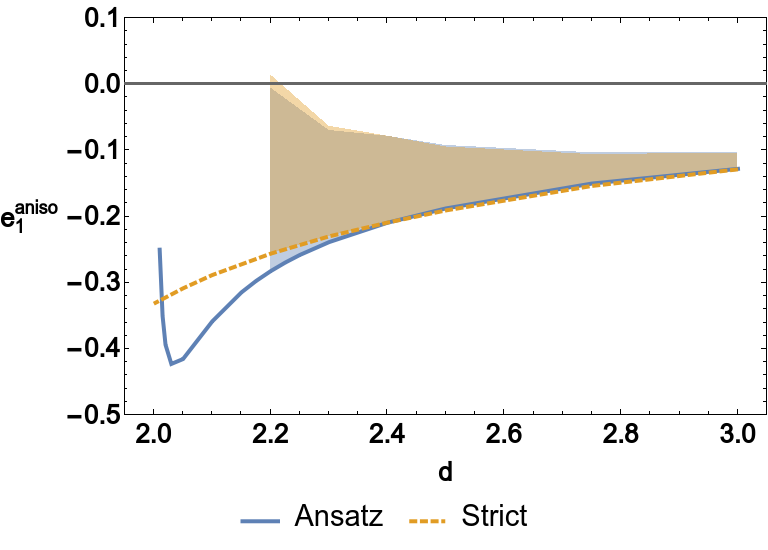}
\caption{(Color online) The dependence of exponent $e_1^{aniso}=y_4$ on the implementation of the DE as a function of $d$. Lines represent raw values from the ansatz and strict versions of the DE2. Marked regions represents DE2 confidence intervals for each implementation.}
\label{y4DEversions}
\end{center}
\end{figure}

We note however that the situation is much less favourable when analyzing dimensions lower than 3. As already discussed in Sect.~\ref{resultslowd} when $d$ decreases the quality of the different approximations performed deteriorates. Several independent indications seem to show this. First, the results of the $\tau$-expanded version $\Gamma_E$ (see Eq.~(\ref{expanded})) begin to differ more and more from those obtained from the one with full field content $\Gamma_F$ (see Eq.~(\ref{double_functional})). In particular, some eigenvalues of the stability matrix cross in the $\Gamma_E$ version and not in the $\Gamma_F$ version. Second, the difference between two consecutive orders of the DE becomes very large. In fact, below $d\sim 2.5$ the value of $y_4$ in the leading order of the DE (the LPA) changes sign, contrary to what is observed at order DE2. This is the origin of the very large confidence intervals of the DE for these dimensions that can be seen in Figs.~\ref{etaDEversions}, \ref{nuDEversions} and \ref{y4DEversions}. In those figures, raw values, marked with lines represents the results coming from ansatz and strict versions of the DE. In Fig.~ \ref{y4DEversions} we employ the $\Gamma_F$ version of the calculation because, as mentioned before, both versions have significant differences for anisotropic eigenvalues below $d\sim 2.5$. In low dimensions, some exponents do not even show an alternating behavior, so we employ in those cases the pessimistic estimate of error bars. 
In this context it is to be expected (and properly found, as shown in Figs.~\ref{etaDEversions}, \ref{nuDEversions} and \ref{y4DEversions}) that the results from the ansatz version and the strict version differ increasingly. These differences should be attributed to higher estimated error bars of the DE in low dimensions. This difficulty becomes extreme as we approach $d\to 2$ where the LPA no longer shows a fixed point. In that case our procedure of calculating the error bars directly becomes meaningless. Indeed, in Figs.~\ref{etaDEversions}, \ref{nuDEversions} and \ref{y4DEversions} the error bars are not represented below $d=2.1$ ($d=2.2$ in anisotropic case) because we loose control of the LPA approximation in those dimensions. Let us note that putting aside the $y_4$ exponent in a small range of dimensions, estimates coming from the strict and the ansatz versions are fully compatible within error bars. Accordingly, the rise of discrepancy is just a manifestation of the fact that higher orders of the DE have a much larger contribution below $d\sim 2.5$.

Despite the facts discussed in the previous paragraph, it should be noted that the results obtained in the ansatz version of DE2 turn out to describe both qualitatively and, to some extent also quantitatively, the expected behavior of the KT transition (including the perturbations with anisotropy $\mathbb{Z}_4$) when $d\to 2$, in contrast to what happens in the strict version. In particular, a sudden drop in the eigenvalue $e_1^{iso}=1/\nu$ around $d=2.2$ is correctly observed in the ansatz version and below this dimension $1/\nu$ approaches zero. This difference in the quality of the results obtained in low dimension may indicate a possible preference among both versions of the DE when approaching dimension 2. This requires further clarifying study. 

Although at the moment we do not have a conclusive explanation, we permit ourselves here to conjecture a possible reason for this substantial difference in quality between the two versions of the DE at $d=2$. As mentioned before, the ansatz version has the property of exactly preserving the generalized response-fluctuation relations, as well as the Ward identities for the $O(2)$ symmetry\footnote{Ward identities and response-fluctuation relations depend only on the existence of an ansatz with symmetry from which all vertices are extracted; it does not require that the ansatz is an accurate description of the physics.}. The strict version, by treating each vertex independently, only satisfies these relations up to higher-order terms in the  momentum expansion. Such relationships are key in the study of the broken phase in the presence of continuous symmetry. Although, strictly speaking, at $d=2$ there is no broken phase in the $O(2)$-symmetric case, at $d=2+\eps$ the broken phase does exist and the transition fixed point approaches the low-temperature fixed point as $\eps\to 0$. In fact, the ansatz version of DE2 is able to exactly describe the behavior at $d=2+\eps$ for $N>2$ at order $\mathcal{O}(\eps)$. Although there is a non-analytical behavior at $d=N=2$ \cite{Chlebicki_2021},
it is expected that this closeness to an exact result improves the quality of the approximation also for $N=2$ in the two-dimensional limit. The resolution of this point calls for an extension of the present study to the DE4 order.

\section{Summary and outlook}
The derivative expansion of the nonperturbative RG has in very recent years been shown to provide a computational tool capable of delivering accurate and highly precise results with controllable error estimates for the universal properties of the $O(N)$ models in $d=3$. In the present paper we have extended this approach to a situation involving a discrete $\mathbb{Z}_4$-symmetric perturbation, dangerously irrelevant at the isotropic $O(2)$-symmetric fixed point. An accurate treatment of such a situation within nonperturbative RG requires substantial methodological advancements as compared to the isotropic case, since the functions parametrizing the flowing effective action depend in an essential way on two field variables. Systematically implementing the DE up to order $\partial^2$, in $d=3$ we have provided an estimate of the leading eigenvalue $y_4$ related to the discrete anisotropy. We have demonstrated that an expansion in the $\mathbb{Z}_4$ invariant $\tau=\frac{1}{2}\phi_1\phi_2$ yields results practically equivalent to those obtained within the complete DE2 approach in that dimension. This is however no longer the case for lower $d$. We have analyzed the dependence of $y_4$ on dimensionality and shown its nonmonotonous character. We finally demonstrated the approach of $y_4$ towards zero for $d\to 2^+$, which marks the onset of nonuniversal critical behavior. Previous NPRG studies, while providing correct resolution of the interplay between the different fixed points and the rich and interesting crossover behavior of the system, did not deliver the accurate eigenvalues describing the cubic perturbation (neither in $d=3$, nor in $d=2$). Our present study demonstrates how this is achieved via a systematic implementation of the derivative expansion. We discussed and compared the different implementations of the DE (the ``ansatz'' and the ``strict'' version). While in the vicinity of $d=3$ these give practically identical results, substantial differences occur in lower $d$ even if the results continue to agree within error bars. In particular, for $d\to 2^+$, only the ``ansatz'' version correctly accounts for the divergence of the correlation length exponent accompanying the onset of the KT physics. This discrepancy calls for further clarifying studies, necessary going beyond the DE2 truncation level. The present work also points to the fertility of the methodology developed recently in Refs.~\onlinecite{Balog_2019, Balog_2020, Polsi_2020, Polsi_2021} in situations reaching beyond the paradigm of isotropic $O(N)$ models. It would be very interesting to extend the present study to the DE4 trucation level. We expect that in $d=3$ this would allow for obtaining the value of $y_4$ with an accuracy better than in previous studies performed so far. On the other hand, in $d=2$ we anticipate that the DE4 calculation might be capable of capturing the KT physics fully accurately and in any case would provide a stringent test of the NPRG error estimate methodology developed recently. This would additionally shed light on the differences between the ansatz and strict variants of the DE approximation scheme.  

%%%%%%%%%%%%%%%%%%%%%%%%%%%%%%%%%%%%%%%%%
\begin{acknowledgments}
We are very grateful to Nicolas Dupuis, Mateusz Homenda and Gonzalo de Polsi for valuable comments on the manuscript. We thank Bertrand Delamotte for numerous discussions and collaboration on related problems. A. Ch. and P. J. acknowledge support from the Polish National Science Center via 2017/26/E/ST3/00211; C. S. and N. W. thank for the support of the Programa de Desarrollo de las Ciencias Básicas (PEDECIBA). The authors acknowledge the Cluster-Uy where some of the numeric calculations have been performed.
\end{acknowledgments}

\bibliography{refs}{} 
%\bibliography{/Users/Pawel/Desktop/Paper_with_Piotr/Version_PJ/refs}{} 
\bibliographystyle{apsrev4-1}

%\bibliography{/Users/piotr/Dropbox/Paper_with_Piotr_2020/Text/Ref}{} 
%\bibliography{/Users/Pawel/Desktop/Paper_with_Piotr_2020/refs}{} 
%\bibliographystyle{apsrev4-1}

\end{document}